\documentclass[%
pre,showpacs,superscriptaddress,
 amsmath,amssymb,
 reprint,%
]{revtex4-1}
\usepackage{amsfonts}

\usepackage{graphicx}
\usepackage{dcolumn}
\usepackage{bm}
\usepackage{amssymb}
\usepackage{array}
\usepackage{longtable}
\usepackage{natbib}

\usepackage{hhline}

\begin{document}

\title{Dynamics of ``comb-of-comb" networks}

\author{Hongxiao Liu}
\author{Yuan Lin}
\affiliation {School of Computer Science, Fudan University, Shanghai
200433, China}
\affiliation {Shanghai Key Laboratory of Intelligent Information
Processing, Fudan University, Shanghai 200433, China}

\author{Maxim Dolgushev}
\email{dolgushev@physik.uni-freiburg.de}
\affiliation {Institute of Physics, University of Freiburg, Hermann-Herder-Str.3, D-79104 Freiburg, Germany}
\affiliation {Institut Charles Sadron, Universit\'e de Strasbourg \& CNRS, 23 rue du Loess, 67034 Strasbourg Cedex,
France}

\author{Zhongzhi Zhang}
\email{zhangzz@fudan.edu.cn}
\affiliation {School of Computer Science, Fudan University, Shanghai
200433, China}
\affiliation {Shanghai Key Laboratory of Intelligent Information
Processing, Fudan University, Shanghai 200433, China}

\begin{abstract}
The dynamics of complex networks, being a current hot topic of many scientific fields, is often coded through the corresponding Laplacian matrix. The spectrum of this matrix carries the main features of the networks' dynamics. Here we consider the deterministic networks which can be viewed as ``comb-of-comb" iterative structures. For their Laplacian spectra we find analytical equations involving Chebyshev polynomials, whose properties allow one to analyze the spectra in deep. Here, in particular, we find that in the infinite size limit the corresponding spectral dimension goes as $d_s\rightarrow2$. The $d_s$ leaves its fingerprint in many dynamical processes, as we exeplarily show by considering the dynamical properties of the polymer networks, including single monomer displacement under a constant force, mechanical relaxation, and fluorescence depolarization.
\end{abstract}
\pacs{36.20.Ey, 36.20.-r, 05.60.Cd, 89.75.Hc, 05.45.Df}

\maketitle
\section{introduction}\label{Introduction}

The interest to the theory of networks shows in the last decade an accelerating growth, by attracting scientists not only from physics but also more and more from the interdisciplinary fields~\cite{Bi15}. Such a transfer of knowledge between different scientific fields is possible because of the mutual underlying mathematics. One of such mathematical fundamental objects is the Laplacian matrix~\cite{CvDoSa98,Ne10}. From the physical point of view, it describes in a very simple way interactions between nearest-neighboring nodes, so that in the case of an infinite linear chain one obtains a discrete form of the Laplacian operator~\cite{DoEd86}.

In macromolecular science the Laplacian matrix is used to reflect the relationship between structural properties of macromolecules and their dynamics~\cite{GuBl05}. Such a concept of macromolecular representation (called generalized Gaussian structures, GGS~\cite{GuBl05}) extends the well-known Rouse model for linear chains~\cite{Ro53} to arbitrary structures. The basic simplicity of the GGS model is that one can obtain analytical solutions of dynamical problems, even for complex polymer systems. Here the deterministic structures are of special interest, because they allow exact calculations typically based on iterative schemes~\cite{RaTo83,CoKa92,JaWuCo92,JaWu94,CaCh97,GoMa02,BlFeJuKo04,GaBl07,Ga10,Ag08,LiWuZh10,JuVoBe11,LiDoQiZh15}. (We note that for disordered networks in some cases mean-field results are possible~\cite{Mo9901,GrGrTi12,GrGrKuTi15,KuDoMu15}.) Having a pool of different structures which carry well-defined, unique spectral properties is very important for checking of general concepts, such as scaling~\cite{ReGrKl08,ReGrKl10,ReKlGr12a,ReKlGr12b,MeChVoBe11,DoGuBlBeVo15}.

In this paper we consider hierarchical ``comb-of-comb" networks. We note that combs, being also synthesized structures~\cite{KoIaLoHa05} (representing up to now, however, only low order iterations of ``comb-of-comb" networks), attract in general, a lot of interest~\cite{AgSaCaCa15,Io12,So12}. Combs are used in context of reaction-diffusion problems~\cite{AgSaCaCa15}, for description of diffusion of ultracold atoms~\cite{Io12} and particles in crowded environments~\cite{So12}, to name only a few examples. The combs which we consider here are hierarchically constructed: at each step to each node of the preceding structure a linear spacer of the same length is attached. We show that the Laplacian spectra of such ``comb-of-comb" networks can be calculated based on the equations involving recursive Chebyshev polynomials, whose properties are of much help for the analysis of the spectra. As an application, we consider relaxation dynamics of the ``comb-of-comb" polymeric networks in the GGS scheme as well as the energy transfer on the networks.

The paper is structured as follows: In Sec. \ref{The model} we introduce the model of ``comb-of-comb" networks  and discuss briefly their structural properties. In Sec.~\ref{Laplacian spectra} we obtain the closed-form formulae for the determination of Laplacian spectra as well as discuss their properties. In Sec.~\ref{GGS and RP} we exemplify the results of Sec.~\ref{Laplacian spectra} on the dynamical properties of macromolecules and on the fluorescence depolarization.  We summarize the conclusions in Sec.~\ref{Conclsions}.

\section{The model}\label{The model}

\begin{figure}[h]
\centering
\includegraphics[width=0.8\linewidth,trim=0 0 0 0]{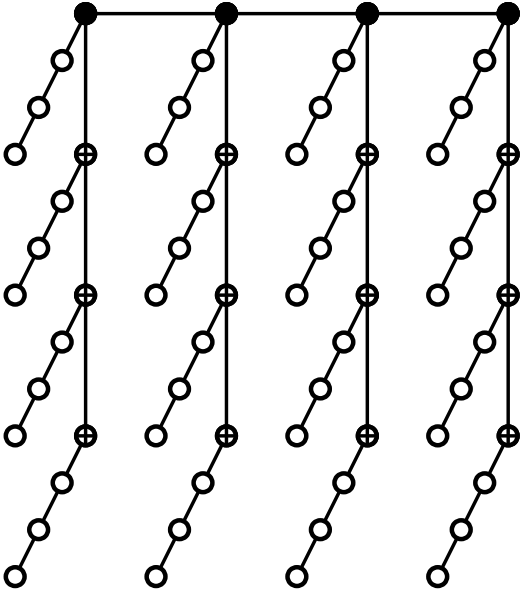}
\caption{Illustration of network $C_g$ corresponding to $r=4$ and $g=3$ (all beads). Only filled beads represent the network at generation $g=1$, filled and crossed beads together give the network of generation $g=2$, see text for details.}\label{Figure Network}
\end{figure}

In this section, we introduce a tree-like network built in an iterative way, which can be viewed as ``comb-of-comb" network. Let $C_g$ ($g \geq 0$) be the family of networks after $g$ iterations. For the initial status $g=0$, $C_0$ is a single node without any edges connected to it. For generation $g=1$ the $C_g$ is just a chain of $r$ nodes, and for $g=2$ it is a comb, see Fig. \ref{Figure Network}. In general, the network $C_{g+1}$ of the generation $g+1$ is obtained by attaching of a new chain with $r-1$ nodes (here $r \geq 2$) to each of the nodes of the network $C_g$, see Fig. \ref{Figure Network}.

By its construction, it is easy to see that the number of nodes and edges in $C_g$ is $N_g=r^g$ and $E_g=r^g-1$, respectively. The tree-like structure assures that the two quantities differ from each other just by $1$. Also there are some other properties: The diameter of this network in generation $g$ is $(2g-1)(r-1)$, which grows logarithmically with the network size, showing that the networks are of small-world type. One can find that the distribution here is not a power law as many other small-world networks, but exponential.  Also the networks $C_g$ have the same number of nodes as, e.g., Vicsek Fractals~\cite{BlFeJuKo04} for $r=f+1$, where $f$ is the only variable in Vicsek Fractals. This makes them quite suitable to compare to other deterministic structures in order to highlight the role of connectivity.

\section{Laplacian spectra}\label{Laplacian spectra}

The Laplacian matrix $\mathbf{L}_g$ is defined through its diagonal elements, which are equal to the degrees (functionalities) of the beads, and through the off-diagonal elements $-1$ for any two directly connected nodes; all other elements of $\mathbf{L}_g$ are zero~\cite{Bi93}.

From the construction of the network $C_g$, we can easily see that the corresponding matrix $\mathbf{L}_g$ is a $r^g\times r^g$ matrix, which can be represented by $r \times r$ blocks:

\begin{equation}
\begin{array}{l}
{\mathbf{L}_g}  = \left( {\begin{array}{*{20}{c}}
{{\mathbf{L}_{g - 1}} + {\mathbf{I}_{g - 1}}}&{ - {\mathbf{I}_{g - 1}}}& \cdots &\textbf{0}&\textbf{0}&\textbf{0}\\
{ - {\mathbf{I}_{g - 1}}}&{2{\mathbf{I}_{g - 1}}}& \cdots &\textbf{0}&\textbf{0}&\textbf{0}\\
\textbf{0}&{ - {\mathbf{I}_{g - 1}}}& \cdots &\textbf{0}&\textbf{0}&\textbf{0}\\
 \vdots & \vdots & \ddots & \vdots & \vdots & \vdots \\
\textbf{0}&\textbf{0}&\cdots &{2{\mathbf{I}_{g - 1}}}&{ - {\mathbf{I}_{g - 1}}}&\textbf{0}\\
\textbf{0}&\textbf{0} & \cdots &{ - {\mathbf{I}_{g - 1}}}&{2{\mathbf{I}_{g - 1}}}&{ - {\mathbf{I}_{g - 1}}}\\
\textbf{0}&\textbf{0}& \cdots &\textbf{0}&{ - {\mathbf{I}_{g - 1}}}&{{\mathbf{I}_{g - 1}}}
\end{array}} \right),
\end{array}
\end{equation}
where each block is a square matrix with size $r^{g-1}$. One can find the eigenvalues of $\mathbf{L}_g$ by solving its characteristic polynomial ${P_g}(\lambda ) = \det (\lambda {{\bf{I}}_g} - {\mathbf{L}_g})$.

To find the $r^g$ roots of the characteristic equation $P_g(\lambda)=0$, we have to make the determinant diagonalization. Here we transform the above determinant into a lower triangle determinant by the procedure discussed below. Let us define $\mathbf{R}_i$ as the $i$th row of block matrices $\lambda {{\bf{I}}_g} - {\mathbf{L}_g}$ in $P_g(\lambda)$ and $\mathbf{D}_i$ as the corresponding diagonal block after diagonalization. The procedure starts from $\mathbf{R}_r$ according to the following steps:

    (1) We add $-\frac{1}{\lambda-1}\mathbf{R}_r$ to $\mathbf{R}_{r-1}$.
        Noting that $\mathbf{D}_r=(\lambda -1)\mathbf{I}_{g-1}$, the diagonal block in row $r-1$ then becomes
        \begin{equation}
        {\mathbf{D}_{r - 1}} = (\lambda  - 2)\mathbf{I}_{g - 1} - \mathbf{D}_r^{-1} = \frac{{\lambda ^2} - 3\lambda  + 1}{\lambda  - 1}{\mathbf{I}_{g - 1}}.
        \end{equation}

    (2) We add $-\mathbf{D}_{r-1}^{-1}\mathbf{R}_{r-1}$ to $\mathbf{R}_{r-2}$,
        the diagonal block in row $r-2$ then follows
        \begin{equation}
        {\mathbf{D}_{r-2}} = (\lambda  - 2){\mathbf{I}_{g - 1}} - \mathbf{D}_{r-1}^{-1} = \frac{{\lambda ^3} - 5{\lambda ^2} + 6\lambda  - 1}{{\lambda ^2} - 3\lambda  + 1}{\mathbf{I}_{g - 1}}.
        \end{equation}

    (3) Analogously, we added $-\mathbf{D}_{i+1}^{-1}\mathbf{R}_{i+1}$ to $\mathbf{R}_i$ ($i=r-3,r-4,...2$),
        the diagonal block in row $i$ then reads
        \begin{equation}
        {\mathbf{D}_{i}} = (\lambda  - 2){\mathbf{I}_{g - 1}} - \mathbf{D}_{i+1}^{-1} = \frac{{(\lambda  - 2){\mathbf{D}_{i+1}} - \mathbf{I}_{g - 1}}}{\mathbf{D}_{i+1}}{\mathbf{I}_{g - 1}}.
        \end{equation}

    (4) Finally, we add $-\mathbf{D}_{2}^{-1}\mathbf{R}_{2}$ to $\mathbf{R}_1$, the element in the upper-left corner becomes
        \begin{equation}
        {\mathbf{D}_{1}} = \frac{(\lambda  - 2){\mathbf{D}_{2}} - \mathbf{I}_{g - 1}}{\mathbf{D}_{2}}{\mathbf{I}_{g - 1}}-{\mathbf{L}_{g - 1}}.
        \end{equation}

For further simplification, we introduce $\alpha_i$ and $\beta_i$ such as $(\alpha_i/\beta_i)\mathbf{I}_{g - 1}=(\lambda  - 2){\mathbf{I}_{g - 1}} - \mathbf{D}_{i+1}^{-1}$. In this notation
\begin{equation}\label{d_i}
\mathbf{D}_{i} = \frac{{{\alpha _i}}}{{{\beta _i}}}{\mathbf{I}_{g - 1}} =  \frac{(\lambda - 2){\alpha _{i + 1}} - {\beta _{i + 1}}}{\alpha _{i + 1}}{\mathbf{I}_{g - 1}}.
\end{equation}
With this we obtain
\begin{equation}
\begin{array}{l}
\begin{array}{*{20}{l}}
{{P_g}(\lambda ) = \det(\mathbf{D}_{1}\mathbf{D}_{2} \cdots \mathbf{D}_{r}) = }\\
{\det\left(\frac{(\lambda  - 2){\mathbf{D}_{2}} - \mathbf{I}_{g - 1}}{\mathbf{D}_{2}}{\mathbf{I}_{g - 1}}-{\mathbf{L}_{g - 1}}\right){{\left(\frac{\alpha _2}{\beta _2}\frac{\alpha _3}{\beta _3} \cdots \frac{\alpha _r}{\beta _r}\right)}^{r^{g - 1}}}}\\
{ = \det\left((\lambda  - 1 - \frac{\beta _2}{\alpha _2}\right){\mathbf{I}_{g - 1}} - {\mathbf{L}_{g - 1}})\alpha _2^{r^{g - 1}}}
\end{array}\\
= {P_{g - 1}}\left(\lambda  - 1 - \frac{\beta _2}{\alpha _2}\right)\alpha _2^{r^{g - 1}}.
\end{array}
\end{equation}

\begin{widetext}

\begin{figure}[h]
\centering
\includegraphics[width=1\linewidth,trim=0 0 0 0]{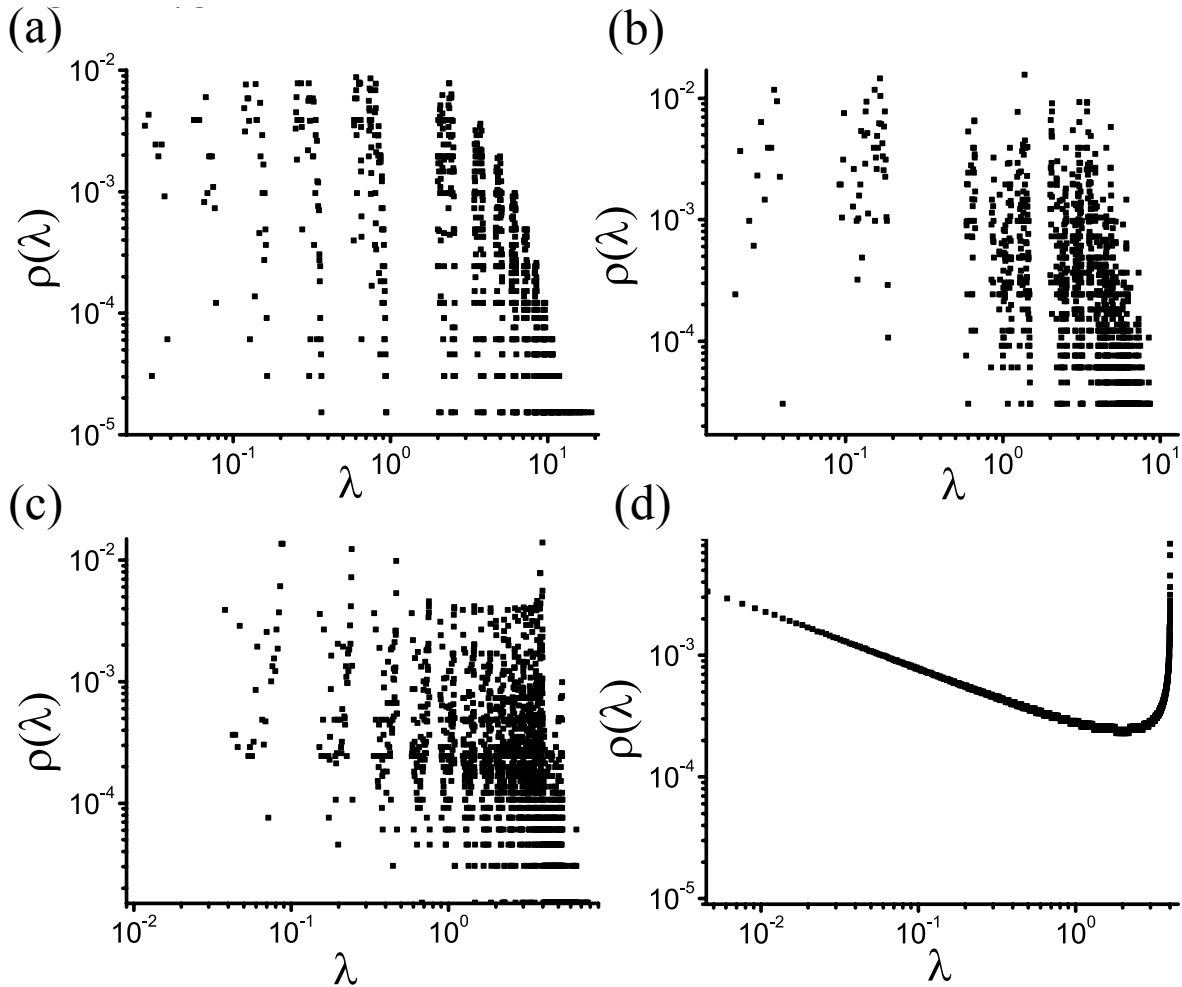}
\caption{Densities of states $\rho(\lambda)$ corresponding to the parameter sets $(r,g)$: (a) $(2,16)$, (b) $(4,8)$, (c) $(16,4)$, (d) $(65536,1)$. All networks have $N_g=65536$ beads in total (case (d) represents just a linear chain), see text for details.}\label{Figure_density}
\end{figure}

\end{widetext}

The factor $\alpha_2$ has an exponent $r^{g-1}$, which is equal to the size of determinant $P_{g-1}(\lambda)$. We can infer that the eigenvalues of $P_g(\lambda)$ are totally determined from those $\lambda^{(g-1)}$ in generation $g-1$ by the equation
\begin{equation}\label{lambda_ab}
\lambda  - 1 - \frac{{\beta _2}}{{\alpha _2}}=\lambda^{(g-1)},
\end{equation}
without the influence of factor $\alpha_2$. It is easy to find the relations between $\alpha$'s and $\beta$'s from Eq.~(\ref{d_i}) as
\begin{equation}\label{alpha_i}
{\alpha _i} = (\lambda  - 2){\alpha _{i + 1}} - {\beta _{i + 1}}\text{ and }  {\beta _i} = {\alpha _{i + 1}}.
\end{equation}
Based on the initial values $\alpha_r=\lambda-1$ and $\beta_r=1$, $\alpha_2$ and $\beta_2$ can be expressed by:
\begin{equation}\label{alpha_2}
{\alpha _2} = W_{r-1}\left(\frac{\lambda}{2}-1\right)\text{ and }  {\beta _2} = W_{r-2}\left(\frac{\lambda}{2}-1\right).
\end{equation}
In Eq.~(\ref{alpha_2}) $W_n(x)$ is the Chebyshev polynomial of the fourth kind~\cite{MaHa03}, for which holds $W_0(x)=1$, ${W_1(x)=2x+1}$, and $W_n(x)=2xW_{n-1}(x)-W_{n-2}(x)$. Moreover, using the relation between the $W_n(x)$ and the Chebyshev polynomial of the second kind $U_n(x)$ (for $U_n(x)$ holds $U_0(x)=1$, $U_1(x)=2x$, and $U_n(x)=2xU_{n-1}(x)-U_{n-2}(x)$): $W_n(x)=U_n(x)+U_{n-1}(x)$, the characteristic Eq.~(\ref{lambda_ab}) becomes
\begin{equation}\label{lambda_g_g-1}
\frac{\lambda U_{r-1}\left(\frac{\lambda}{2}-1\right)}{W_{r-1}\left(\frac{\lambda}{2}-1\right)} = {\lambda^{(g - 1)}},
\end{equation}
where the only variable here is the eigenvalue $\lambda$, corresponding to generation $g$. Thus, one can obtain all eigenvalues iteratively, starting from the exact eigenvalues of the discrete linear chain\cite{Py68} of length $r$, $\lambda_k^{(1)}=4\sin^2[(k-1)\pi/(2r)]$ for $k=1,\dots,r$.

Now, recalling that the substitution $x=\cos\theta$ leads to $U_n(x)=\sin(n+1)\theta/\sin\theta$ and to ${W_n(x)=\sin(n+\frac{1}{2})\theta/\sin\frac{1}{2}\theta}$ or that $x=\cosh\theta$ leads to ${U_n(x)=\sinh(n+1)\theta/\sinh\theta}$ and to ${W_n(x)=\sinh(n+\frac{1}{2})\theta/\sinh\frac{1}{2}\theta}$~\cite{MaHa03}, Eq.~(\ref{lambda_g_g-1}) can be readily solved. Moreover, with the substitution $x=\cos\theta$ we can look at the behavior of small eigenvalues. Making a Taylor expansion of the left-hand side of Eq.~(\ref{lambda_g_g-1}), we obtain:
 \begin{equation}
r\lambda +\mathcal{O}(\lambda ^2)\approx{\lambda^{(g - 1)}} \text{ for }\lambda\ll1.
 \end{equation}
Thus, given that the total number of nodes $N_g$ at generation $g$ is related to that of $g-1$ by $N_g=rN_{g-1}$, the density of states $\rho(\lambda)$ follows $\rho(\lambda)=\rho(\lambda/d)$. Hence the spectral dimension $d_s$, which is defined through~\cite{AlOr82}
\begin{equation}
\rho(\lambda)\sim\lambda^{d_s/2-1},
\end{equation}
readily follows for the $C_g$ networks, $d_s=2$. As we proceed to show the $d_s$, being a key parameter for the dynamical characteristics, leaves its fingerprints in their behavior. In Fig.~\ref{Figure_density} we exemplify the density of states $\rho(\lambda)$ of $C_g$ for different parameter sets $(r,g)$. As can be inferred from the figure, for higher $g$ the density of states tends to the $\rho(\lambda)\sim\lambda^0$ behavior. Figure~\ref{Figure_density}(d) reproduces the well-known~\cite{AlOr82} result for linear chains $\rho(\lambda)\sim\lambda^{-1/2}$.

\section{Relaxation dynamics and fluorescence depolarization}\label{GGS and RP}

In this section, we illustrate the dynamical behavior of ``comb-of-comb" networks $C_g$ in the GGS formalism~\cite{SoBl95,Sc98,GuBl05}, which  extends the Rouse model~\cite{Ro53} (for linear chains) to complex architectures. The macromolecules in such a framework are represented by beads connected by springs, just as the nodes connected by edges in networks. Each bead is located at time $t$ at the position vector $\mathbf{R}_i(t)$, $i=1,\ldots,N_g$. Each of them experiences friction with friction constant $\zeta$ and the springs which connect them have elasticity constant $K$, see review~\cite{GuBl05} for details.

The dynamics in GGS follows a set of Langevin equations~\cite{GuBl05}:
\begin{equation}\label{Langevin}
\zeta \frac{d \mathbf{R}_m(t)}{dt} + K\, \sum_{i=1}^{N_g} L_{mi}\mathbf{R}_m(t) = \mathbf{f}_m(t) + \mathbf{F}_m(t)\,.
\end{equation}
In Eq.~(\ref{Langevin}), $L_{mi}$ is the $mi$th entry of the Laplacian matrix $\mathbf{L}_g$ describing the topology of the network, which has been introduced in Sec.~\ref{Laplacian spectra}; $\mathbf{f}_m(t)$ is the thermal noise, assumed to be Gaussian with $\langle \mathbf{f}_m(t) \rangle=0$ and $\langle f_{m \alpha}(t)f_{n \beta}(t')\rangle=2k_B T\delta_{\alpha \beta}\delta_{nm}\delta(t-t')$, where $k_B$ is the Boltzmann constant, $T$ is the temperature, $\alpha$ and $\beta$ denote the directions in three-dimensions. $\mathbf{F}_m(t)$ denotes another (possible) external force acting on $m$th bead.

We focus on the motion of the GGS under a constant external force $\mathbf{F}_m = F\theta(t-0)\delta_{mk}\mathbf{e}_y$, switched on at $t=0$ and acting on a single bead (say, $k$th) in the $y$ direction.  After averaging over the random forces $\mathbf{f}_m(t)$ and over all the beads in the GGS, the displacement~\cite{SoBl95,Sc98,BiKaBl00} along this direction is given by
\begin{equation}\label{Y(t)}
\langle Y(t) \rangle = \frac{F t}{N_g \zeta} + \frac{F}{\sigma N_g \zeta} \sum_{i=2}^{N_g} \frac{1 -\exp(-\sigma \lambda_i^{(g)} t)}{\lambda_i^{(g)}}\,,
\end{equation}
where $\sigma=K/\zeta$ is the bond rate constant. The $\lambda_i^{(g)}$s are the eigenvalues of  $\mathbf{L}_g$, except the eigenvalue $0$ denoted by $\lambda_1^{(g)}$, related to the displacement of the center of mass. We note that the motion of a specific bead does also depend on the eigenvectors of $\mathbf{L}_g$, see Ref.~\cite{BiKaBl00} for details; picking the bead randomly and performing ensemble averaging leads to the simple form of Eq.~(\ref{Y(t)})~\cite{SoBl95,Sc98,BiKaBl00}.

\begin{figure}[h]
\centering
\includegraphics[width=1\linewidth,trim=0 0 0 0]{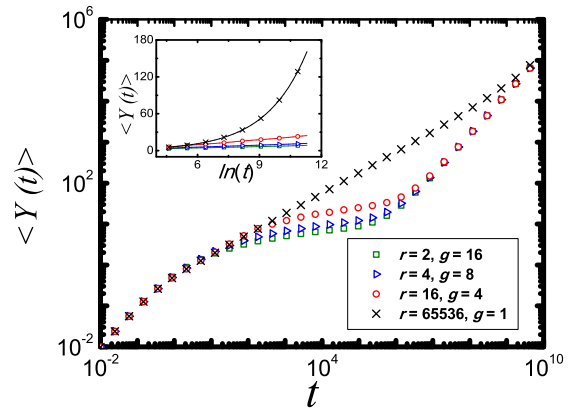}
\caption{(Color online) Averaged monomer displacement $\left\langle {Y(t)} \right\rangle$, Eq.~(\ref{Y(t)}), for the networks $C_g$ of different $r$ and $g$. The case $r=2^{16}=65536$ and $g=1$ corresponds to a linear chain. All networks have $N_g=65536$ beads in total. The choice of units is given by setting $\sigma=1$ and $\frac{F}{\zeta}=1$, see text for details.}\label{Figure Y(t)}
\end{figure}

The global properties of $\langle Y(t) \rangle$ can be extrapolated by Eq.~(\ref{Y(t)}): its behavior for very short times is $\langle Y(t) \rangle \sim F t/ \zeta$ while for very long times, we have $\langle Y(t) \rangle \sim F t /(N_g \zeta)$. These are general features for all the GGS structures. The particular GGS architecture is revealed only in the intermediate time domain, on which we concentrate in Fig. \ref{Figure Y(t)}. The $\langle Y(t) \rangle$ of $C_g$ shows in this domain a logarithmic behavior, meaning that the the network's beads move very slowly before the whole network starts the diffusive motion. We note that such a behavior differs from that of the typical fractal structures, such as the Vicsek fractals, for which
$\langle Y(t) \rangle \sim t ^{\gamma}$ (with $0<\gamma<0.5$)~\cite{BlFeJuKo04}, as well as from the linear chains for which $\langle Y(t) \rangle \sim t ^{1/2}$. Logarithmic behavior of $\langle Y(t)\rangle$ is observed, however, for dendrimers~\cite{BiKaBl00,BiKaBl01}. Nevertheless, as we proceed to show, the mechanical relaxation of ''comb-of-comb" networks $C_g$ differs qualitatively from that of the dendrimers.

In the mechanical relaxation experiments one measures the response to harmonically applied external forces. The result is the complex dynamic modulus $G^*(\omega)$, in other words, the $G'(\omega)$ and $G''(\omega)$ (storage and loss modulus)~\cite{DoEd86,Fe80},
\begin{equation}\label{G'}
G'(\omega ) = \frac{\nu k_B T}{N_g}\sum_{i=2}^{N_g} {\frac{{{{(\omega /2\sigma {\lambda_i^{(g)}})}^2}}}{{1 + {{(\omega /2\sigma {\lambda_i^{(g)}})}^2}}}}
   \end{equation}
and
\begin{equation}\label{G''}
G''(\omega ) = \frac{\nu k_BT}{N_g}\sum\limits_{i = 2}^{N_g} {\frac{{\omega /2\sigma {\lambda_i^{(g)}}}}{{1 + {{(\omega /2\sigma {\lambda_i^{(g)}})}^2}}}}\,.
\end{equation}
In Eqs.~(\ref{G'})-(\ref{G''}), $\nu$ denotes the number of polymer monomers (beads) per unit volume.

\begin{figure}[h]
\centering
\includegraphics[width=1\linewidth,trim=0 0 0 0]{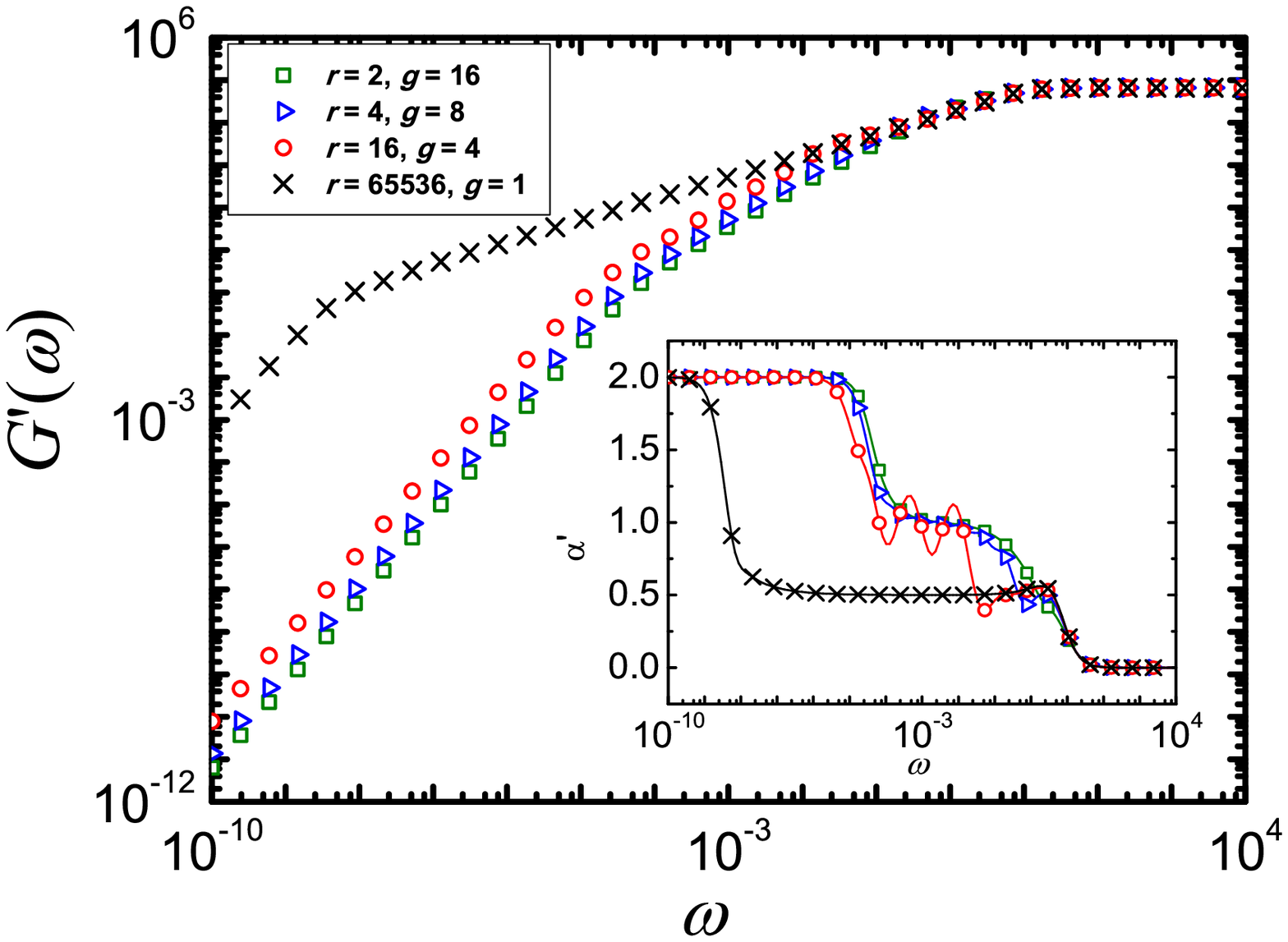}
\includegraphics[width=1\linewidth,trim=0 0 0 0]{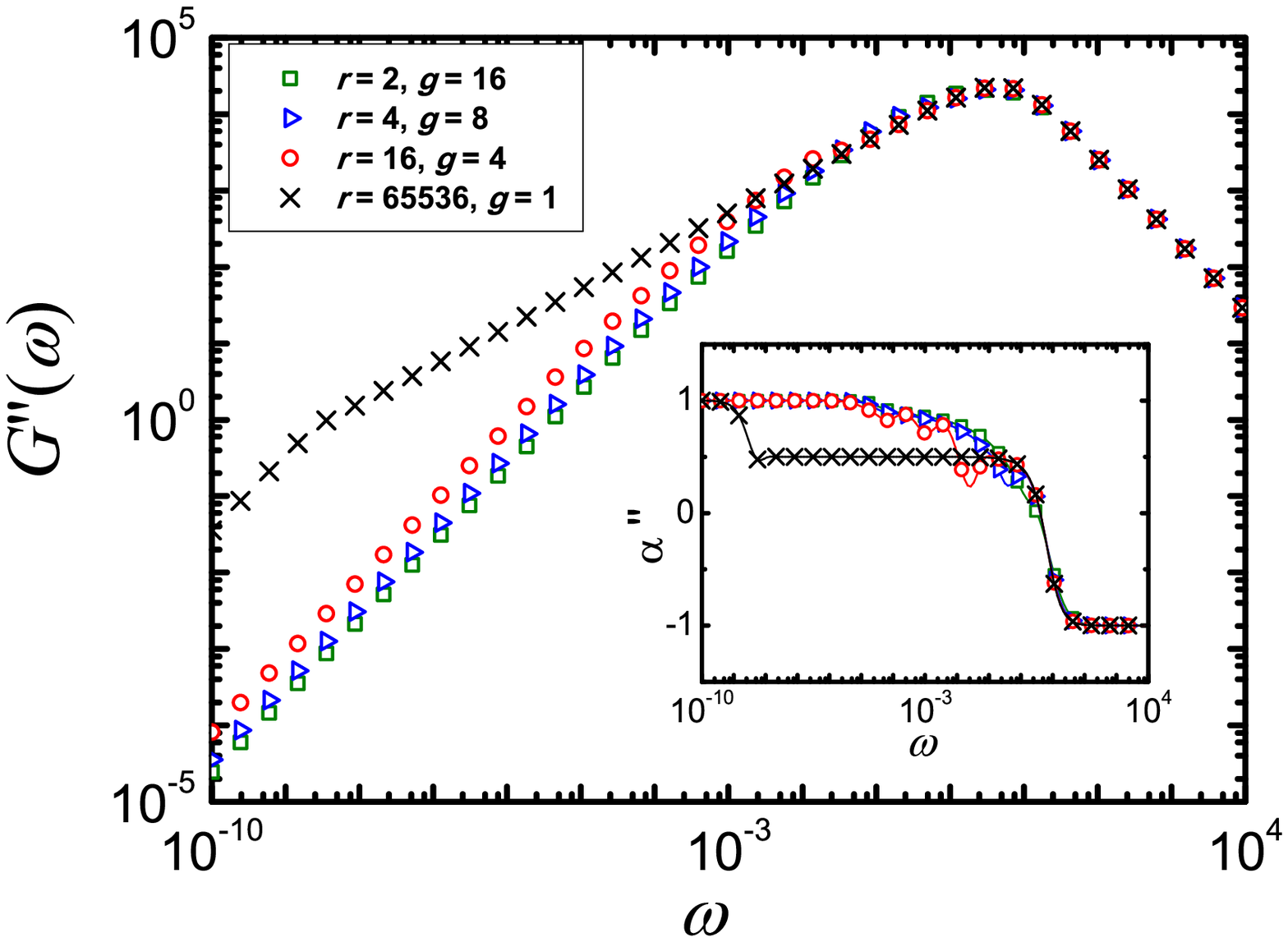}
\caption{(Color online) Top figure: storage modulus $G'(\omega)$, Eq.~(\ref{G'}); bottom figure: loss modulus $G''(\omega)$, Eq.~(\ref{G''}); both for the networks of Fig.~\ref{Figure Y(t)}. The inset represents local slopes of the curves with the same symbolic and color code.  Here we use $\sigma=1$ and $\frac{\nu k_BT}{N_g}=1$, see text for details.}\label{Figure G'(t)}
\end{figure}

Again, as for $\langle Y(t)\rangle$, the moduli at low and high frequencies are independent of structure. Here we focus on storage and loss moduli, for which one has $G'(\omega)\sim \omega^2$ and $G''(\omega)\sim \omega^1$ at very small $\omega$ and $G'(\omega) \sim \omega^0$ and $G''(\omega)\sim \omega^{-1}$ at very high $\omega$. The inbetween region of $G'(\omega)$ (corresponding to the intermediate times in $\langle Y(t) \rangle$) shows for $C_g$-networks in double-logarithmic scales a slope with the exponent around $1$, see Fig. \ref{Figure G'(t)}. The $G''(\omega)$ shows a continuous transition between $1$ and $-1$ slopes. For a better visualization, we plot the effective slopes $\alpha ' = \frac{{d({{\log }_{10}}G')}}{{d({{\log }_{10}}\omega )}}$ or $\alpha'' = \frac{{d({{\log }_{10}}G'')}}{{d({{\log }_{10}}\omega )}}$ for $G'(\omega)$ or $G''(\omega)$ of the main plots of Fig. \ref{Figure G'(t)} as insets to them. As Fig. \ref{Figure G'(t)} shows, for very low and very high frequencies, the limiting behaviors of $\alpha'$ or $\alpha''$ yield the slopes $2$ or $1$ and $0$ or $-1$, respectively.  The wavy behavior of $\alpha'$ or $\alpha''$ in the intermediate frequency region is due to the high symmetry of $C_g$-networks. A similar behavior has been observed for many other structures~\cite{BlFeJuKo04,JuFrBl02,LiDoQiZh15}, it reflects the high regularity of the system. Typical fractals, such as Vicsek fractals, lead to a fractional slope between $1/2$ and $1$~\cite{BlFeJuKo04}, as also observed experimentally for (disordered) hyperbranched polymers~\cite{SoKlBl02}. (However, for dendrimers, which are not fractals, $G'(\omega)$ shows in the intermediate frequency domain a logarithmic behavior~\cite{BiKaBl00,BiKaBl01}.) The linear chains follow a $G'(\omega)\sim \omega^{1/2}$ behavior~\cite{DoEd86} (see also the black curve on Fig. \ref{Figure G'(t)}).

\begin{figure}[h]
\centering
\includegraphics[width=1\linewidth,trim=0 0 0 0]{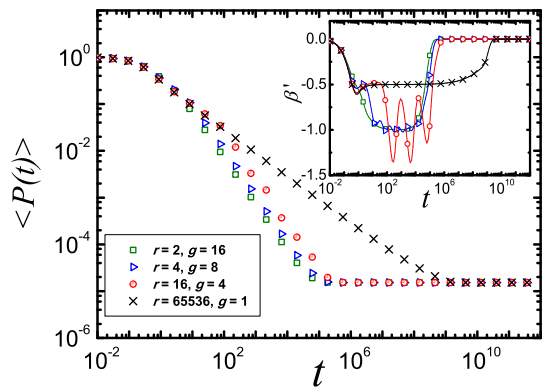}
\caption{(Color online) The average probability $\langle P(t) \rangle$, Eq.~(\ref{Energy transfer(3)}), for the networks of Fig.~\ref{Figure Y(t)}. The inset represents local slopes of the curves with the same symbolic and color code. The time is given in units of $1/\tilde{k}$ by setting $\tilde{k}=1$, see text for details.}\label{Energy transfer 1}
\end{figure}

The Laplacian spectra are important not only for the dynamics \textit{of} polymeric networks, but also for the dynamics \textit{on} networks. A classical example is the energy transfer over a system of chromophores~\cite{BlVoJuKo0501,BlVoJuKo0502,GaBl07,BaKlKo97,BaKl98}. As a usual way, we suppose that the energy can be directly transferred only between the nearest neighbors of each node. Under these conditions the dipolar quasiresonant energy transfer among the chromophores can be investigated by the following equation~\cite{BlVoJuKo0501,BlVoJuKo0502,GaBl07}
\begin{equation}\label{Energy transfer(1)}
\frac{dP_i(t)}{dt}= \sum^{N_g}_{\substack{j = 1\\ j \ne i}} T_{ij}P_j(t)-\left(\sum^{N_g}_{\substack{j = 1\\ j \ne i}} T_{ij}\right)P_i(t)\,,
\end{equation}
where $P_i(t)$ means the probability that node $i$ is excited at time $t$ and $T_{ij}$ denotes the transfer rate from node $j$ to node $i$.

As in~\cite{BlVoJuKo0501,BlVoJuKo0502,GaBl07}, we separate the radiative decay, which is equal for all chromophores, from the transfer problem. The radiative decay leads only to the multiplication of all the $P_i(t)$ by $\exp(-g/\tau_R)$, where $1/\tau_R$ is the radiative decay rate. With the assumption that all microscopic rates are equal to each other, say $\tilde{k}$, Eq.~(\ref{Energy transfer(1)}) becomes
\begin{equation}\label{Energy transfer(2)}
\frac{dP_i(t)}{dt} =-\tilde{k}\,\sum^{N_g}_{\substack{j = 1\\ j \ne i}} L_{ij}P_j(t)- \left(\tilde{k} L_{ii}\right)P_i(t)\,,
\end{equation}
where $L_{ij}$ is the $ij$th entry of Laplacian matrix $\mathbf{L}_g$. Note in Eq.~(\ref{Energy transfer(2)}), the relation $L_{ii}=-\sum_{j\neq i}L_{ji}$ holds. By averaging over all sites, the probability of finding the excitation at time $t$ on the originally excited chromophore depends only on the eigenvalues (and not on the eigenvectors) of $\mathbf{L}_g$ and is given by\cite{BlVoJuKo0501,BlVoJuKo0502,GaBl07}
\begin{equation}\label{Energy transfer(3)}
\langle P(t) \rangle  = \frac{1}{N_g}\sum_{i = 1}^{N_g} \exp (-\tilde{k}\, \lambda_i^{(g)}\,t)\,.
\end{equation}

 In Fig. \ref{Energy transfer 1} we display the results of the average probability $\langle P(t) \rangle$ that an initially excited chromophore of $C_g$ is still or again excited at time $t$. In the intermediate time domain (most of the differences appear here) the decays obey a power-law behavior as $\langle P(t) \rangle \sim t^{-\beta'}$. In Fig. \ref{Energy transfer 1} the $\beta'$ oscillates around $1$ after $t=10^2$ (see also the local slope in the inset obtained from the corresponding derivative). Thus, the decay is faster than that for linear chains (for them $\langle P(t) \rangle \sim t^{-1/2}$ holds\cite{AlOr82}) and than that for typical fractals (for Vicsek fractals of functionality $f=3$ and $f=4$ one has $\beta'\approx0.56$ and $\beta'\approx0.59$, respectively\cite{BlVoJuKo0501,BlVoJuKo0502}). Also there is a qualitative difference to dendrimers, for which no scalings for $\langle P(t) \rangle$ in the intermediate time domain are observable~\cite{BlVoJuKo0501}.

\section{Conclusions}\label{Conclsions}

Laplacian matrix is a one of the most important objects describing interactions in many-component systems, such as networks. Here we have studied the Laplacian spectra of the deterministic structures, which can be viewed as ``comb-of-comb" networks. We found that the spectra can be determined recursively from an analytical equation, which involves Chebyshev polynomials. The knowledge of properties of the Chebyshev polynomials allowed us to determine the related spectral dimension $d_s$.

Here we have illustrated the importance of these findings for polymeric networks. In particular, we looked at the (micro)rheological properties by considering motion of a monomer under applied constant force as well as by investigating mechanical relaxation moduli. The dynamics on the networks was illustrated on the dipolar quasiresonant energy transfer. In all considered quantities the spectral dimension $d_s$ plays a fundamental role.

We note that our findings will be interesting not only for polymers, but also for many other fields, e.g., for quantum walks~\cite{KuDoMu15,MuBl11} and for mean-first passage problems~\cite{BeGuVo15,BeVo14}, as well as in general for network theory~\cite{Bi15,Ne10,multiplex,AlDi07}.

\begin{acknowledgements}
This work was supported by the National Natural Science Foundation of China under Grant No. 11275049. M.D. acknowledges DFG through IRTG ``Soft Matter Science" (GRK 1642/1).
\end{acknowledgements}



\begin{thebibliography}{9}
\bibitem{Bi15}
G. Bianconi,
EPL {\bf 111}, 56001 (2015).
%
\bibitem{CvDoSa98}
D. M. Cvetkovi\'{c}, M. Doob, and H. Sachs. \textit{Spectra of Graphs: Theory and Applications} (Wiley, New York, 1998).
%
\bibitem{Ne10}
M. Newman. \textit{Networks: An Introduction} (Oxford University Press, 2010)
%
\bibitem{DoEd86}
M. Doi and S. F. Edwards, \emph{The Theory of Polymer Dynamics} (Clarendon, Oxford, 1986).
%
\bibitem{GuBl05}
A. A. Gurtovenko and A. Blumen,
Adv. Polym. Sci. {\bf 182}, 171 (2005).

\bibitem{Ro53}
P. E. Rouse,
J. Chem. Phys. {\bf 21}, 1272 (1953).




\bibitem{RaTo83}
R. Rammal and G. Toulouse, J. Phys. Lett. \textbf{44}, 13 (1983).

\bibitem{CoKa92}
M. G. Cosenza and R. Kapral, Phys. Rev. A \textbf{46}, 1850 (1992).

\bibitem{JuFrBl02}
A. Jurjiu, C. Friedrich, and A. Blumen,
Chem. Phys. \textbf{284}, 221 (2002).


\bibitem{JaWuCo92}
C. S. Jayanthi, S. Y. Wu, and J. Cocks,
Phys. Rev. Lett. {\bf 69}, 1955 (1992).

\bibitem{JaWu94}
C. S. Jayanthi and S. Y. Wu,
Phys. Rev. B {\bf 50}, 897 (1994).

\bibitem{BlFeJuKo04}
A. Blumen, C. von Ferber, A. Jurjiu, and T. Koslowski,
Macromolecules {\bf 37}, 638 (2004).

\bibitem{CaCh97}
C. Cai and Z. Y. Chen,
Macromolecules {\bf 30}, 5104 (1997).

\bibitem{GoMa02}
Y. Y. Gotlib and D. A. Markelov, Polym. Sci. Ser. A \textbf{44}, 1341 (2002).

\bibitem{GaBl07}
M. Galiceanu and A. Blumen,
J. Chem. Phys. {\bf 127}, 134904 (2007).

\bibitem{Ga10}
M. Galiceanu, J. Phys. A \textbf{43}, 305002 (2010).

\bibitem{Ag08}
E. Agliari, Phys. Rev. E \textbf{77}, 011128 (2008).

\bibitem{LiWuZh10}
Y. Lin, B. Wu, and Z. Z. Zhang
Phys. Rev. E \textbf{82}, 031140 (2010).

\bibitem{JuVoBe11}
 A. Jurjiu, A. Volta, and T. Beu, Phys. Rev. E {\bf 84}, 011801 (2011).

\bibitem{LiDoQiZh15}
H. X. Liu, M. Dolgushev, Y. Qi, Z. Z. Zhang. Sci. Rep. \textbf{5}, 9024 (2015).



\bibitem{Mo9901}
R. Monasson,
Eur. Phys. J. B {\bf 12}, 555-567 (1999).

\bibitem{GrGrTi12}
C. Grabow, S. Grosskinsky, and M. Timme,
Phys. Rev. Lett. {\bf 108}, 218701 (2012).

\bibitem{GrGrKuTi15}
C. Grabow, S. Grosskinsky, J. Kurths, and M. Timme,
Phys. Rev. E {\bf 91}, 052815 (2015).


\bibitem{KuDoMu15}
N. Kulvelis, M. Dolgushev, and O. M\"ulken,
Phys. Rev. Lett. {\bf 115}, 120602 (2015).


\bibitem{ReGrKl08}
S. Reuveni, R. Granek, and J. Klafter,
Phys. Rev. Lett. {\bf 100}, 208101 (2008).
\bibitem{ReGrKl10}
S. Reuveni, R. Granek, and J. Klafter,
Proc. Natl. Acad. Sci. USA {\bf 107}, 13696 (2010).
\bibitem{ReKlGr12a}
S. Reuveni, J. Klafter, and R. Granek,
Phys. Rev. Lett. {\bf 108}, 068101 (2012).
\bibitem{ReKlGr12b}
S. Reuveni, J. Klafter, and R. Granek,
Phys. Rev. E {\bf 85}, 011906  (2012).

\bibitem{MeChVoBe11}
B. Meyer, C. Chevalier, R. Voituriez, and O. B\'{e}nichou, Phys. Rev. E {\bf 83}, 051116 (2011).

\bibitem{DoGuBlBeVo15}
M. Dolgushev, T. Gu\'{e}rin, A. Blumen, O. B\'{e}nichou, and R. Voituriez, Phys. Rev. Lett. {\bf 115}, 208301 (2015).

\bibitem{KoIaLoHa05}
G. Koutalas, H. Iatrou, D. J. Lohse, and N. Hadjichristidis,
Macromolecules  {\bf 38}, 4996-5001 (2005).

\bibitem{AgSaCaCa15}
E. Agliari, F. Sartori, L. Cattivelli, and D. Cassi,
Phys. Rev. E {\bf 91}, 052132 (2015).

\bibitem{Io12}
A. Iomin,
Phys. Rev. E {\bf 86}, 032101 (2012).

\bibitem{So12}
I. M. Sokolov,
Soft Matter {\bf 8}, 9043 (2012).

\bibitem{Bi93}
N. Biggs,
\textit{Algebraic Graph Theory}, 2nd ed. (Cambridge University Press, Cambridge, England, 1993).

\bibitem{MaHa03}
J. C. Mason and D. C. Handscomb,
\textit{Chebyshev Polynomials} (Chapman and Hall, London, 2003).

\bibitem{Py68}
C. W. Pyun,  J. Chem. Phys. \textbf{49}, 2875 (1968).

\bibitem{AlOr82}
S. Alexander and R. Orbach, J. Phys. Lett. \textbf{43}, 625 (1982).

\bibitem{SoBl95}
J. U. Sommer and A. Blumen,
J. Phys. A {\bf 28}, 6669 (1995).

\bibitem{Sc98}
H. Schiessel,
Phys. Rev. E {\bf 57}, 5775 (1998).

\bibitem{BiKaBl00}
P. Biswas, R. Kant, and A. Blumen,
Macromol. Theory Simul. {\bf 9}, 56 (2000).

\bibitem{BiKaBl01}
P. Biswas, R. Kant, and A. Blumen,
J. Chem. Phys. {\bf 114}, 2430 (2001).

\bibitem{Fe80}
J. D. Ferry,
\emph{Viscoelastic Properties of Polymers}, 3rd ed. (Wiley, New York, 1980).

\bibitem{SoKlBl02}
\bibinfo{author}{I. M. Sokolov}, \bibinfo{author}{J. Klafter}, and  \bibinfo{author}{A. Blumen},
\newblock \bibinfo{journal}{Physics Today}
  \textbf{\bibinfo{volume}{55}}, \bibinfo{issue}{11}, \bibinfo{pages}{48--54}
  (\bibinfo{year}{2002}).

\bibitem{BlVoJuKo0501}
A. Blumen, A. Volta, A. Jurjiu, and T. Koslowski,
J. Lumin. {\bf 111}, 327 (2005).

\bibitem{BlVoJuKo0502}
A. Blumen, A. Volta, A. Jurjiu, and T. Koslowski,
Physica A {\bf 356}, 12 (2005).

\bibitem{BaKlKo97}
A. Bar-Haim, J. Klafter, and R. Kopelman,
J. Am. Chem. Soc. {\bf 119}, 6197 (1997).

\bibitem{BaKl98}
A. Bar-Haim and J. Klafter,
J. Phys. Chem. B {\bf 102}, 1662 (1998).


\bibitem{MuBl11}
O. M\"ulken and A. Blumen, Phys. Rep. {\bf 502}, 37 (2011).

\bibitem{BeVo14}
O. B\'{e}nichou and R. Voituriez, Phys. Rep. {\bf 539}, 225 (2014).

\bibitem{BeGuVo15}
O. B\'{e}nichou,  T. Gu\'{e}rin, and R. Voituriez, J. Phys. A {\bf 48}, 163001 (2015).


\bibitem{multiplex}
S. Boccaletti, G. Bianconi, R. Criado, C.I. del Genio, J. G\'{o}mez-Gardenesi, M. Romance, I. Sendina-Nadal, Z. Wang, and M. Zanin,
Phys. Rep. {\bf 544}, 1 (2014).

\bibitem{AlDi07}
J. A. Almendral and A. D\'{i}az-Guilera,
New J. Phys. {\bf 9}, 187 (2007).


\end{thebibliography}
\end{document}